\documentclass{moriond}

\def\be{\begin{equation}}
\def\ee{\end{equation}}
\def\bea{\begin{eqnarray}}
\def\eea{\end{eqnarray}}

\newcommand{\Photo}{}

\begin{document}
\vspace*{4cm}
\title{Homogeneity in the search of the Dark Sector}

\author{ Pierros Ntelis and Anne Ealet} 

\address{Aix Marseille Univ, CNRS/IN2P3, CPPM, Marseille, France}
\maketitle

\abstracts{ In the era of precision cosmology, we investigate an novel method to probe the Dark Universe. By studying the fractality of the universe, we estimate a characteristic scale of homogeneity.  Using a fisher analysis, we investigate the potential impact of this scale to the cosmological information we can gain from the Euclid satellite, and therefore understand the nature of Dark Energy and Dark matter of our universe. 
}

\section{Introduction}
{\small The $\Lambda$CDM model is the most predominant model describing the Universe today and it is based on the Cosmological Principle. This Principle states that the universe is homogeneous and isotropic on scales large enough. The purpose of our research is to estimate the power of Euclid to measure the scale at which the universe becomes homogeneous. This method was investigated in a recent thesis framework\cite{PIERROS_THESIS}. Euclid is a promising future satellite that is going to map the 3D structure of our universe and allows us to perform such measurement. }
\subsection{Euclid Mission}
It is a satellite project\cite{Euclid} that is going to give us first light in 2021. The mission will use an $1.2$ SiC mirror telescope that is going to fly in the Sun-Earth L2 Lagrange point for 6 years. 
It will perform:
\begin{itemize}\setlength\itemsep{0em}
\item Imaging with the $VIS$ instrument at optical regime: $550 < \lambda/nm < 900$
\item Photometry with the $NISP$ Instrument at $Y,J,H$ bands: $900 < \lambda/nm < 2000$
\item Slittless Spectroscopy with $NISP$ with:
	\begin{itemize} \setlength\itemsep{0em}
		\item $R=380$ (low resolution, fast reduction)
		\item $920 < \lambda/nm < 1850$
		\item $30$ Million Targets per $4000$ sec
		\item  Cover an area of $15000$ $deg^2$ 
	\end{itemize} 
	
\item Main Objectives:
	\begin{itemize}\setlength\itemsep{0em}
	\item  Large scale structure science
	\item  Baryon Acoustic Oscillations
	\item  Weak Gravitational Lensing
	\item  Ameliorate SuperNovae Measurements
	\end{itemize}
\end{itemize}
To probe Dark Matter, Dark Energy and the growth of structures in the late universe.

\section{Methodology}

Our observable is the Fractal Correlation Dimension as a function of scales, $D_2(r)$. This quantity is directly related to the average number of galaxies within a cell of a spherical volume, $N(r)$, via the equation:
\begin{equation}
	D_2(r) = \frac{d \ln N(r)}{d \ln r}
\end{equation}
We get that for $D_2<3$, an inhomogeneous distribution, $D_2=3$, a homogeneous and for $D_2>3$, a super-homogeneous distribution.
The 1\% homogeneity defines a characteristic scale:
\begin{equation}
	D_2(R_H) = 2.97\; .
\end{equation}
To produce the 3D galaxy map from the observational data, i.e. comoving positions of galaxies, we need to convert the redshift, z, Right Ascension, R.A. and Declination, DEC, into cartesian comoving coordinates according to a flat $\Lambda$CDM model: 
\begin{equation}
	\chi(z) = \frac{c}{H_0} \int^z_0 \frac{dz'}{(\Omega_b + \Omega_{cdm})(1+z')^3+\Omega_{\Lambda}}
\end{equation}
with the Planck 2015 fiducial cosmology
\begin{equation}
	(\Omega_m,\Omega_{\Lambda},\Omega_{b},h,n_s,\ln\left[ 10^{10}A_s\right]) = (0.32,0.68,0.01,0.67,0.97,3.09)
\end{equation}

\subsection{Empirical Error}
In order to see how the error on the Homogeneity Scale scales with future experiments we vary the density $n$, volume, $V$ and photometric redshift error, $\sigma_z$ of $100$ qpm mock catalogues that follow the $BOSS$ $CMASS$ galaxy sample\cite{ntelis2017exploring}. We find that the Precision of the Fractal Correlation Dimension depends only on the Volume according to the empirical formula:
\begin{equation}
	\sigma_{D_2}^{\mathrm{emp}} = (0.0063\mp 0.0021) \sqrt{\frac{2.16h^{-3}\mathrm{Gpc}^3}{V}} D_2
\end{equation}
This estimate is conservatively $30\%$ accurate in the scales $[50-120]\ \mathrm{Mpc}/h$ where we measure the homogeneity scale.

\subsection{Study Dark Matter and Dark Energy}
To see the cosmological impact of this probe, we perform a Fisher Analysis Forecast to estimate the sensitivity of the Homogeneity Scale against the ratio density of Cold Dark Matter, $\omega_{cdm}$, the Dark Energy ratio density, $\Omega_{\Lambda}$ , the dimensionless Hubble constant, $h$ and the equation of state parametrised by $w_0$. In particular we model the two cosmological probes, the characteristic scale of homogeneity, $R_H$, and its competitor, and usual probe, $R_{BAO}$, according to the following fisher formalism:
\begin{equation}
	F_{ij} =  \sum_{z=0}^{4} \frac{1}{\sigma_{o}^{2}(z)} \frac{\partial o(z) }{\partial p_{i}} \frac{\partial o(z) }{\partial p_{j}}
\end{equation}
where o(z) is the observable at each redshift bin, $o(z) = \left\{ R_H(z), R_{BAO}(z) \right\} $ and $\sigma_{o}(z)$ is their correspoding error. Notice that we have summed up the 5 redshift bins of our in study galaxy sample. 

\begin{figure}

\begin{minipage}{1.0\linewidth}
\centerline{\includegraphics[width=0.9\linewidth]{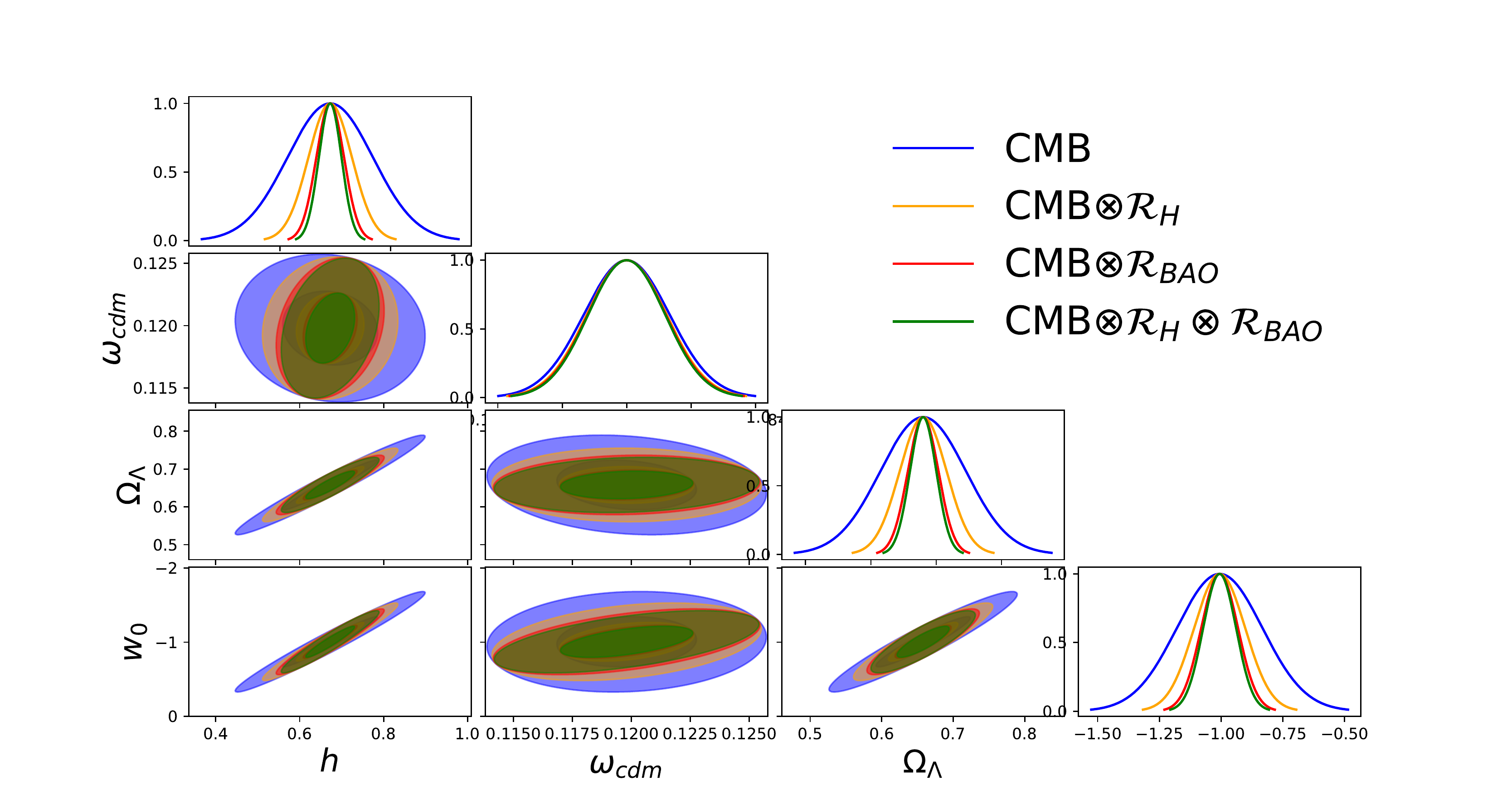}}
\end{minipage}

\caption[]{Fisher $1\sigma$ and $2\sigma$ contours prediction of different probes combined with the Cosmic Microwave Background (CMB). $R_H$ is the homogeneity scale and $R_{BAO}$ is the Baryon Acoustic Oscillations scale.}
\label{fig:Results2}
\end{figure}

The diagram \ref{fig:Results2}  shows the $1\sigma$ and $2\sigma$ contours of the aforementioned cosmological parameters using the latest Cosmic Microwave Background (CMB) measurements, and those combined with the Homogeneity scale and the BAO scale. Clearly the homogeneity scale provides an extra information, that will help us investigate the Alternative Cosmological Scenarios. 

\section{Conclusions}
We expect that the universe has a fractal-like behaviour at small scales due to clustering and behaves as a homogeneous fluid at large scales as previous studies have shown\cite{PIERROS_THESIS} using the fractality of the universe. We have investigate the error of the homogeneity scale probe as we expect it from future experiments such the one of Euclid mission Satellite, which is going to give first light in 2021. We have shown with a Fisher Analysis that the homogeneity scale is a complementary cosmological probe and it will help us understand the nature of Dark Energy and Dark Matter of our universe.

\section*{Acknowledgments}

{\small We would like to thank the Moriond scientific comittee that select us to present our research in the form of poster in this wonderful environment. Special thanks to JC Hamilton, JM Le Goff, J rich, N Busca, S. Escoffier, A. Tilquin, and W. Gillard. PN would like to express his sincere gratitude to Peter Schneider on the reduction of this research. The computations were performed in the scientific computer MARDEC of IN2P3/CNRS. PN is supported by CNES. }

\end{document}